# INDUCED GRAVITY IN THE SHORT RANGE

C.P. Kouropoulos[†]

## *Abstract*

We consider a pair of harmonic oscillators in two or three dimensions of space coupled by the standard electrodynamic forces : the Coulomb, the Lorentz and the electrokinetic forces. The addition of the Lorentz force is mainly felt in the short range and suppresses the radial correlated oscillating mode of such coupled oscillators. This imposes constraints on the system that make the two transverse modes degenerate. As a result, an *1/r* antigravitational interaction now appears in the surviving anticorrelated radial zero-mode, which does not allow coherent states to form. As gravitation can only emerge from coherent modes, it can no longer be transitive. Matter in high densities would thus tend to increase its disorder, decouple from its own gravity, from the ordering far infrared Machian background that coheres its rest energy and would become intrinsically unstable. The highly energetic jets from galactic nuclei could be the consequence.

General Physics

[†]email : kouros@bluewin.ch

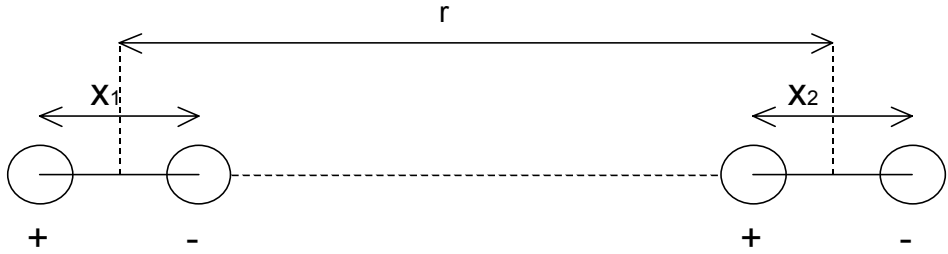

This represents two electric dipoles. Their mutual Coulomb coupling is

$$[1] \quad 4\pi\varepsilon_o U_{12} = -\frac{e^2}{r+\frac{x_1+x_2}{2}} - \frac{e^2}{r-\frac{x_1+x_2}{2}} + \frac{e^2}{r+\frac{x_1-x_2}{2}} + \frac{e^2}{r-\frac{x_1-x_2}{2}}$$

$$= \frac{e^2}{r}\left(-\frac{1}{1+\frac{x_1+x_2}{2r}} - \frac{1}{1-\frac{x_1+x_2}{2r}} + \frac{1}{1+\frac{x_2-x_1}{2r}} + \frac{1}{1-\frac{x_2-x_1}{2r}}\right),$$

$$\frac{1}{1+\frac{a}{r}} = 1 - \frac{a}{r} + \frac{a^2}{r^2} - \ldots \quad ,$$

$$U_{12} = -\frac{1}{2\pi\varepsilon_o}\frac{e^2}{r^3}x_1 x_2 + \ldots$$

$$F_1 = -\partial_{x_1} U_{12} = \frac{1}{2\pi\varepsilon_o}\frac{e^2}{r^3}x_2$$

Besides the Coulomb coupling, let us consider the velocity and acceleration-dependent forces from the Lagrangian

[2] $L = \frac{1}{2}m(\dot{x}_1^2 + \dot{x}_2^2) + \frac{m\omega_o^2}{2}(\vec{x}_1^2 + \vec{x}_2^2) - U_{12} + \dot{\vec{x}}_i \vec{A}_{ij}$

with the forces

[3] $\vec{F}_i = \partial_{\bar{x}_i} L - d_t \frac{\partial U}{\partial \dot{\bar{x}}_i} = -f \vec{x}_i + \frac{1}{2\pi\varepsilon_o} \frac{e^2}{r_o^3} x_j - \frac{\mu_o}{2\pi} \frac{e^2}{r} \left( \ddot{\vec{x}}_j - \frac{1}{r}(\dot{\vec{x}}_j \times \hat{r}) \times \dot{\vec{x}}_i \right)$ ,

where we assumed that in each oscillator, the acceleration of one charge opposes that of the other. The equations of motion are

[4] $m\ddot{\vec{x}}_i + f \vec{x}_i - k(\hat{r} \bullet \vec{x}_j)\hat{r} + k' \left( \ddot{\vec{x}}_j - \frac{1}{r}(\dot{\vec{x}}_j \times \hat{r}) \times \dot{\vec{x}}_i \right) = 0$

$k = \frac{1}{2\pi\varepsilon_o} \frac{e^2}{r^3}$ , $k' = \frac{\mu_o}{2\pi} \frac{e^2}{r}$

The term $f$ is akin to a spring's tension and approximates the electrodynamic interaction between the neighbouring charges of one oscillator on a small displacement. Assuming that $r$ is along the $x$ axis and confining ourselves to the $x$-$y$ plane, [4] becomes

[5]  i) $m\ddot{x}_1 + f x_1 - kx_2 + k'\left( \ddot{x}_2 - \frac{1}{r} \dot{y}_2 \dot{y}_1 \right) = 0$

ii) $m\ddot{x}_2 + f x_2 - kx_1 + k'\left( \ddot{x}_1 - \frac{1}{r} \dot{y}_2 \dot{y}_1 \right) = 0$

iii) $m\ddot{y}_1 + f y_1 + k'\left( \ddot{y}_2 + \frac{1}{r} \dot{y}_2 \dot{x}_1 \right) = 0$

iv) $m\ddot{y}_2 + f y_2 + k'\left( \ddot{y}_1 + \frac{1}{r} \dot{y}_1 \dot{x}_2 \right) = 0$

Taking

$$[6] \quad x_1 = Ae^{i\omega t} \quad , \quad x_2 = Be^{i\omega t} \quad , \quad y_1 = Ce^{i\omega t} \quad , \quad y_2 = De^{i\omega t} \quad ,$$

substituting and simplifying by $e^{i\omega t}$, we find that this system has solutions only if the quadratic mixed terms can be eliminated. Radially, *i)-ii)* has a trivial correlated solution $A = B$ that does not allow for oscillation in [5] and a nontrivial oscillating anticorrelated solution $A = -B$. Inserting the former into *iii)* ± *iv)* requires the transverse system to be trivial and non-oscillating too. On the other hand, the radial anticorrelated solution $A = -B$ yields a nontrivial transversely correlated solution to *iii)+iv)* and a nontrivial anticorrelated solution to *iii)-iv)*. Substituting

$$[7] \quad \eta = \frac{1}{\sqrt{2}}(x_1 - x_2) \quad , \quad \upsilon_\pm = \frac{1}{\sqrt{2}}(y_1 \pm y_2) \quad ,$$

into the action yields

$$[8] \quad \begin{aligned} &i) - ii) \quad (m + k')\ddot{\eta} + (f - k)\eta = 0 \\ &iii) \pm iv) \quad (m + k')\ddot{\upsilon}_\pm + f\upsilon_\pm = 0 \quad . \end{aligned}$$

The corresponding eigenfrequencies are

$$[9] \quad \omega_{\eta+} = \omega_o \sqrt{\frac{1 - \frac{k}{m\omega_o^2}}{1 + \frac{k'}{m}}} \cong \omega_o \sqrt{1 - \frac{1}{m}\left(k' + \frac{k}{\omega_o^2}\right)} \quad ,$$

$$\omega_{\eta-} = \{\emptyset\} \quad ,$$

$$\omega_{\upsilon_\pm} = \omega_\upsilon = \frac{\omega_o}{\sqrt{1 + \frac{k'}{m}}} \cong \omega_o \sqrt{1 - \frac{k'}{m}} \quad ,$$

$$\omega_o = \sqrt{\frac{f}{m}} \quad .$$

These can be expanded as

$$[10] \quad \omega_{\eta_+} \cong \omega_o\left(1 - \frac{\Phi_1}{2} - \frac{\Phi_2}{8} - \frac{\Phi_3}{16} - \ldots\right) ,$$

$$\Phi_1 = \psi(\mu_o + \vartheta) ,$$

$$\Phi_2 = \psi^2(3\mu_o^2 + 2\vartheta\mu_o - \vartheta^2) ,$$

$$\Phi_3 = \psi^3(\vartheta^3 - \vartheta^2\mu_o + 3\vartheta\mu_o^2 + 5\mu_o^3) ,$$

$$\psi = \frac{e^2}{4\pi m\, r} , \quad \vartheta = \frac{1}{\varepsilon_o \omega_o^2 r^2} .$$

and

$$[11] \quad \omega_\upsilon \cong \omega_o\left(1 - \frac{\vartheta}{2} - \frac{\vartheta^2}{8} - \frac{\vartheta^3}{16} - \ldots\right) ,$$

$$\vartheta = \frac{e^2 \mu_o}{4\pi m\, r} .$$

Let us write the Hamiltonian corresponding to the equations of motion [8]:

$$[12] \quad H = \frac{1}{2m'}\left(\vec{p}_{\eta_+}^{\,2} + \vec{p}_{\upsilon_+}^{\,2} + \vec{p}_{\upsilon_-}^{\,2}\right) + \frac{m}{2}\left(\omega_{\eta_+}^2 \eta^2 + \omega_{\upsilon_+}^2 \upsilon_+^2 + \omega_{\upsilon_-}^2 \upsilon_-^2\right) ,$$

$$m' = m + k' .$$

We now solve the problem of our coupled oscillators in the light of Planck's and Post's ensemble interpretation of the zero-point energy. Planck's original result involved counting the number of uncoupled harmonic oscillators that partake in the field's motions in a disordered system averaged over some random phase $\phi$ of the oscillators at a given frequency. If instead of phase-independent non-interacting oscillators we have a system of coupled oscillators of which we count the individual correlated motions, this approach yields very different results than for the uncoupled case studied by Planck. This point has also been stressed by Post. We assume that our oscillators can be diagonalized into two fundamental modes, of correlated or anticorrelated motions whose eigenfrequencies are $\omega_-$ and $\omega_+$, that a vast majority of the oscillators are phase-correlated and contribute to the $\omega_-$ mode, that an oscillator whose phase shifts from $\phi$ to $\phi+\pi$ increases the occupation number of one mode while decreasing that of the other, and that one such isolated transition does not substantially affect the two fundamental frequencies of the system.

$$[13] \quad \eta_+ = \hat{q}_{\eta_+} \sin(\omega_{\eta_+} t + \phi_{\eta_+}) \quad , \quad \upsilon_\pm = \hat{q}_\pm \sin(\omega_\pm t + \phi_\pm) \quad .$$

The energy of a single coupling mode is

$$[14] \quad E_{\eta_+} = \frac{1}{2} k_{\eta_+} \hat{q}_{\eta_+}^2 \quad , \quad E_{\upsilon_\pm} = \frac{1}{2} k_{\upsilon_\pm} \hat{q}_{\upsilon_\pm}^2 \quad ,$$

and that of a system bound between condition *I* and condition *II* is found by substituting [14] into [12], and then integrating over the collective phase-space. For the variable $\upsilon$

$$[15] \quad \langle E \rangle = \frac{\iint\limits_{I \to II} (E_{\upsilon_+} + E_{\upsilon_-}) dp_{\upsilon_+} d\upsilon_+ dp_{\upsilon_-} d\upsilon_-}{\iint\limits_{I \to II} dp_{\upsilon_+} d\upsilon_+ dp_{\upsilon_-} d\upsilon_-} \quad ,$$

$$\Lambda(q_\pm, p_\pm) = (\hat{q}_\pm, \phi_\pm) \quad , \quad J(\Lambda) = m\hat{q}_\pm \omega_\pm \quad ,$$

$$\langle E \rangle = \frac{k_+}{4} \frac{\hat{q}_+^{II4} - \hat{q}_+^{I4}}{\hat{q}_+^{II2} - \hat{q}_+^{I2}} + \frac{k_-}{4} \frac{\hat{q}_-^{II4} - \hat{q}_-^{I4}}{\hat{q}_-^{II2} - \hat{q}_-^{I2}}$$

$$= \frac{k_+}{4}\left(\hat{q}_+^{I2} + \hat{q}_+^{II2}\right) + \frac{k_+}{4}\left(\hat{q}_-^{I2} + \hat{q}_-^{II2}\right)$$

$$= \frac{1}{2}(E_I + E_{II})$$

Next, we write the energies corresponding to the conditions *I* and *II* of the system. If the system can be partitioned into two mutually exclusive sets, with a dominant majority in the correlated mode :

$$[16] \quad E_{I-} = \frac{k_-}{2}\hat{q}_-^{I2} = n_-\hbar\omega_- \quad , \quad E_{II-} = (n_- + 1)\hbar\omega_-$$

$$E_{I+} = \frac{k_+}{2}\hat{q}_+^{I2} = n_+\hbar\omega_+ \quad , \quad E_{II+} = (n_+ - 1)\hbar\omega_+$$

$$\langle E \rangle = \frac{1}{2}(E_I + E_{II}) = \hbar\omega_\upsilon\left(n_- + \frac{1}{2}\right) + \hbar\omega_\upsilon\left(n_+ - \frac{1}{2}\right) \quad .$$

$$N = n_+ + n_-$$

The correlation tends to be transitive, even considering retardation. The correlated modes can form coherent stationary or propagating states in large ensembles whereas the anticorrelation tends to self-destruct as the anticorrelated modes mutually exclude one another. So it is natural that the anticorrelating zero-modes should resemble those of a fermionic oscillator and have the corresponding statistics. Supersymmetry naturally emerges in this context, from coupled harmonic oscillators. If a fraction $\kappa$ of the oscillators is in a random uncoupled phase, this fraction will be described by the habitual Bose statistics. As a result, using [11], the mean energy of the transverse modes can be approximated as

$$[17] \quad \langle E \rangle_T \cong 2\hbar\omega_o\left(\kappa\left(n_{T_o} + \frac{1}{2}\right) + (1-\kappa)(n_{\vartheta_+} + n_{\vartheta_-})\left(1 - \vartheta - \frac{\vartheta^2}{2} - \frac{\vartheta^3}{8} - \ldots\right)\right) ,$$

whose only zero mode, which in this case creates antigravity, is from the disordered and decoupled phase. The mean radial energy, however, is

$$[18] \quad \langle E \rangle_R \cong \hbar\omega_o\left(\kappa\left(n_{R_o} + \frac{1}{2}\right) + (1-\kappa)\left(n_{\eta_+} - \frac{1}{2}\right)\left(1 - \Phi_1 - \frac{\Phi_2}{2} - \frac{\Phi_3}{8} - \ldots\right)\right) ,$$

because the correlated radial mode is absent, being forbidden from oscillating. There is now a negative zero-mode. In the event of a quasi continuous spectrum whose upper bound is the near-field, the mean energy being a *1/r* dependence in the zero-modes may only result if $\rho_\omega$ is strongly peaked in the far infrared from [18] in $\hbar(1-\kappa)<\omega_o>\Phi_1/2$ from the lone fermionic radial zero-mode. Not only is it repulsive, but its anticorrelated Fermi statistics forbids it from forming coherent states. Note how the radial zero-mode also contains a series of repulsive Van der Waals terms at higher orders. The addition of the magnetic coupling in two or three dimensions therefore changes the situation relative to one single dimension or to the radiative gauge, kills the correlated radial mode and makes the frequencies of the two transverse excited modes degenerate. But this phenomenon only affects particles in their near mutual field in its accepted sense. If we assume this to be the particle's Compton length, it means that within that range, the particles cannot gravitate nor partake together in global coherent modes as discussed in a previous paper. The mean energy from the zero-modes related to the three dimensions at some frequency is whose antigravitational contribution vanishes with κ. Seing as no gravitating order may arise, it is expected that κ will grow. When κ is small ($0 \leq \kappa < \kappa < \frac{1}{3}(1-\kappa)$), there is gravity, albeit in a disorganizing form, which increases κ. At a given temperature $T$,

$$[21] \quad \rho = Z^{-1} e^{-\frac{H}{kT}} \quad , \quad Z = Tr\, e^{-\frac{H}{kT}} \quad , \quad Z = Z_+ Z_- Z_o \quad ,$$

$$Z_\pm = \sum_{n_\pm=0}^{\infty} \left\langle \varphi_{n_\pm} \left| e^{-\frac{H_\pm}{kT}} \right| \varphi_{n_\pm} \right\rangle = \sum_{n_\pm=0}^{\infty} e^{-\left(n_\pm \mp \frac{1}{2}\right)\frac{\hbar\omega_\pm}{kT}} = \frac{e^{\pm\frac{\hbar\omega_\pm}{2kT}}}{1 - e^{-\frac{\hbar\omega_\pm}{kT}}} \quad .$$

[19]
$$\langle E \rangle = \frac{\int_{\omega=o}^{c/r} \rho(\omega)\langle E \rangle_\omega}{\int_{\omega=o}^{c/r} \rho(\omega)} ,$$

[20]
$$\langle E \rangle_\omega \cong \hbar\langle\omega_o\rangle\left(\frac{3\kappa}{2} - \frac{1-\kappa}{2}\left(1 - \Phi_1 - \frac{\Phi_2}{2} - \frac{\Phi_3}{8} - \ldots\right)\right) ,$$

The $|\phi_{n_\pm}>$ are built from the Hermite-Tschebychev polynomials $H_{n_\pm}$, which allow for normalized orthogonal eigenfunctions in the Hilbert space. The energy is

$$[22] \quad \langle H \rangle_\omega = Tr(H\rho) = Z^{-1} Tr\left(He^{-\frac{H}{kT}}\right) = kT^2 \frac{1}{Z}\frac{dZ}{dT} \quad ,$$

$$= \hbar(1-\kappa)\left(-\frac{\omega_{\eta_+}}{2} + \frac{\omega_{\eta_+}}{e^{\frac{\hbar\omega_{\eta_+}}{kT}}+1} + \frac{2\omega_\upsilon}{\sinh\frac{\hbar\omega_\upsilon}{kT}}\right) + 3\hbar\kappa\omega_o \left(\frac{1}{2} + \frac{1}{e^{\frac{\hbar\omega_o}{kT}}-1}\right)$$

The limits of which are

$$[23] \quad \langle H \rangle = \hbar(1-\kappa)\left(-\frac{\omega_{\eta_+}}{2} + \frac{kT\omega_{\eta_+}}{\hbar\omega_{\eta_+}+2kT} + \frac{2kT}{\hbar}\right) + 3\hbar\kappa\omega_o \left(\frac{1}{2} + \frac{kT}{\hbar\omega_o}\right)$$

$$\cong 2kT + \kappa kT \quad , \quad kT \gg \hbar\omega \quad ,$$

$$= \hbar(1-\kappa)\left(-\frac{\omega_{\eta_+}}{2} + \omega_{\eta_+} e^{-\frac{\hbar\omega_{\eta_+}}{kT}} + 4\omega_\upsilon e^{-\frac{\hbar\omega_\upsilon}{kT}}\right) + 3\hbar\kappa\omega_o \left(\frac{1}{2} + e^{-\frac{\hbar\omega_o}{kT}}\right)$$

$$\cong -\frac{\hbar(1-\kappa)\omega_{\eta_+}}{2} + \frac{3\hbar\kappa\omega_o}{2} \quad , \quad kT \ll \hbar\omega$$

so that, assuming that $\kappa < \tfrac{1}{3}(1-\kappa)$ at moderate temperatures $T$, the attractive but chaos-inducing $\eta_+$ zero-mode still dominates. The excited states have a predominantly positive energy and result in antigravitation. With an increasing temperature, $\kappa$ is expected to increase as the thermal random motions destroy the order, which also governs the zero-modes of the radial anticorrelated states. Such an effect is expected to dominate during the primal inflationary phases of the Universe or within very dense cosmic objects. Each interacting mode extends about the oscillator in some finite near-field volume that shrinks as the cube of the frequency, so that we are not faced with the prospect of integrating the ZPE of empty space, except in the far infrared from distant matter. Such a derivation is well suited to models such as A.O. Barut's unified electromagnetic theory of matter, to Dyonic schemes and may even be applied to the Standard Model. In the very short range, since the Van der Waals forces can be predominantly repulsive, it is expected that it is the electrodynamic interactions, such as from the magnetic moments of spinning matter or the Ampère forces between relativistic currents that take over and hold the particles together. Far away, the magnetic as well as the Coulomb dipole-dipole interactions become negligible before the electrokinetic near-field, and the corresponding terms can be omitted, yielding results similar to our previous analysis.

To conclude, the magnetic interaction induces chaos at short range and destroys the coherence of the mutually gravitating, stable forms of matter, including with the modes of the far Machian background.

References :

M. Planck ; *Theory of Heat Radiation*, (1912)

E.J. Post ; *Quantum Reprogramming*, Kluwer (1995)

A. Sakharov ; *Vacuum Quantum Fluctuations in Curved Space and the Theory of Gravitation*, Soviet Physics Doklady Vol 12, No11p.1040 (1968)

A.O. Barut ; Found. of Physics 20, 1233 (1990)

B. Haisch, A. Rueda & H. Puthoff ; *Advances in the proposed Electromagnetic Zero-Point Field Theory of Inertia*, Physics/9807023 (1998)

T. Boyer ; *Classical electromagnetic interaction of a point charge and a Magnetic Moment,* Physics/0107006

C.P. Kouropoulos ; *The Origin of Gravity,* Physics/0107015